\documentclass[journal,dvips]{IEEEtran}
\usepackage{amsfonts}
\usepackage{cite}
\usepackage[cmex10]{amsmath}

\newcommand{\bs}{\boldsymbol}
\newcommand{\tx}{\textrm}

\newcommand\MYhyperrefoptions{bookmarks=true,bookmarksnumbered=true,
pdfpagemode={UseOutlines},plainpages=false,pdfpagelabels=true,
colorlinks=true,linkcolor={black},citecolor={black},
urlcolor={black},pdftitle={Full 3D Quantum Transport Simulation of Atomistic Interface Roughness in
Silicon Nanowire FETs},
pdfsubject={Interface roughness in nanowire FETs},
pdfauthor={SungGeun Kim, Abhijeet Paul, Mathieu Luisier, Timothy B. Boykin and Gerhard
Klimeck},
pdfkeywords={Gate-all-around (GAA), MOSFET, silicon nanowire (SiNW), interface roughness
(IR), surface roughness (SR), scattering, effective mobility}}
\ifCLASSINFOpdf
\usepackage[\MYhyperrefoptions,pdftex]{hyperref}
\else
\usepackage[\MYhyperrefoptions,breaklinks=true,dvips]{hyperref}
\usepackage{breakurl}
\fi

\input colordvi

   \usepackage[dvips]{graphicx}
   \graphicspath{{./}}
\ifCLASSOPTIONcompsoc
\usepackage[tight,normalsize,sf,SF]{subfigure}
\else
\usepackage[tight,footnotesize]{subfigure}
\fi

\begin{document}

\title{Full 3D Quantum Transport Simulation of Atomistic Interface Roughness in Silicon Nanowire FETs\thanks{%
This work has been supported by NSF grant EEC-0228390 that funds the Network for Computational Nanotechnology, by NSF PetaApps grant number OCI-0749140, and by the Nanoelectronics Research Initiative (NRI) through the Midwest Institute for Nanoelectronics Discovery. The authors acknowledge the support of the MSD Focus Center, one of six research centers funded under the Focus Center Research Program (FCRP), a Semiconductor Research Corporation entity, computational resources from nanoHUB, and following supercomputer resources: Coates and Steele (RCAC), Kraken (NICS), Jaguar (NCCS), and Ranger (TACC).}}
\author{~SungGeun Kim,\thanks{%
SungGeun Kim, Abhijeet Paul, Mathieu Luisier, and Gerhard Klimeck are with
Network for Computational Nanotechnology, School of Electrical and Computer
Engineering, Purdue University, West Lafayette, IN 47907, USA; email:
kim568@purdue.edu}~Abhijeet Paul, \textit{Student Member, IEEE},~Mathieu Luisier,~Timothy B. Boykin\thanks{%
Timothy B. Boykin is with Electrical and Computer Engineering Dept., The
University of Alabama in Huntsville, Huntsville, AL 35899, USA; e-mail:
boykin@ece.uah.edu}, \textit{Senior Member, IEEE}, and~Gerhard Klimeck,
\textit{Senior Member, IEEE}}
\maketitle

\begin{abstract}
 The  influence of interface roughness scattering (IRS)  on the performances of silicon nanowire field-effect transistors (NWFETs) is numerically investigated using a full 3D quantum transport simulator based on the atomistic $sp^3d^5s^*$ tight-binding model. The interface between the silicon and the silicon dioxide  layers is generated in a real-space atomistic representation using an experimentally derived autocovariance function (ACVF). The  oxide layer is modeled  in the virtual crystal approximation (VCA) using fictitious SiO$_2$ atoms. $\left<110\right>$-oriented nanowires with different diameters and randomly generated surface configurations are studied. The experimentally observed ON-current and the threshold voltage is quantitatively captured by the simulation model. The mobility reduction due to IRS is studied through a qualitative comparison of the simulation results with the experimental results.
\end{abstract}


\section{Introduction}

\IEEEPARstart{A}%
s the dimensions of the planar metal-oxide-semiconductor field effect transistor (MOSFET) are approaching  the nanometer scale, their performances become limited by the short channel effects (SCE), an increasing OFF-state current, and a poor electrostatic control of the channel through a single-gate contact.

  The nanowire (NW) field-effect transistor has been identified as a promising candidate to overcome these issues and build the next generation switch \cite{ITRS2009},\cite{Appen2008}.
 Several experimental studies have shown that NWFETs have excellent electrical characteristics\cite{wang2003,Persson2004,Suk2008}, especially in a gate-all-around (GAA) configuration that provides a superior channel electrostatic control \cite{Singh2006,Suk2005}.
  Among other advantages, GAA NWFETs exhibit the smallest SCE as compared to other multi-gate (MG) structures \cite{Park2002,Li2005}. Despite the better electron transport properties of III-V materials, Si has remained the material of choice to fabricate GAA NWFETs due to its well-understood characteristics and its compatibility to the conventional CMOS technology \cite{Cui2001,Yang2004}.

In line with experimental efforts, simulation and modeling approaches have been developed
to help design better performing NWFETs and understand their  physical behavior.
 Until very recently, drift-diffusion (DD) and Monte Carlo (MC)  approaches have been successfully used to simulate large devices. However, at the nanometer scale it is necessary to go beyond DD and MC models and use a quantum mechanical approach that captures the wave nature of electrons\cite{Ren2000}.

Nano-scale devices are characterized by 3D material variations at the atomistic scale so that an atomistic description of the simulation domain has become indispensable. Non-parabolicities and anisotropies of the bandstructure strongly affects the transport characteristics of electrons in NW transistors \cite{Phytos2008Elec,Wang2005jul}. Furthermore, the strong confinement of electrons in nano-devices lifts degeneracies due to valley splitting \cite{Phytos2008}.
These effects cannot be captured by the effective mass approximation. Therefore, it is inevitable to go beyond the effective mass approximation and to use a full-band approach such as the nearest-neighbor empirical tight-binding (ETB) model \cite{Luisier2006}.

 With the recent development of OMEN \cite{Luisier2008}, a 3D full-band quantum transport simulator based on
 the atomistic $sp^3d^5s^*$ TB model \cite{Boykin2004},
 it  has been possible to explore nano-scale NWFETs. However, the computational burden of such an approach is so intense that large computing resources and an efficient parallelization scheme of the work load is required \cite{Luisier2008Para}. 

The focus of this paper is on interface roughness scattering (IRS) that is, in conjunction with electron-phonon \cite{Luisier2009} and impurity scattering \cite{Martinez2009}, crucial to understand the electrical characteristics of ultra-scaled NWFETs \cite{Suk2007}. At the nanometer scale, the details of the NW surface become more important \cite{cui2003,gold2005}. In the ballistic transport limit, the normalized ON-current of NWFETs is expected to increase as the diameter decreases. Experimental data shows that the ON-current increases with decreasing diameter until 3 nm \cite{Suk2007}. However, decreasing the diameter below 3 nm reverses the trend and reduces the current density. This behavior has been attributed to electron-phonon scattering and interface roughness scattering \cite{Suk2007}. Here we examine this experimental reasoning through modeling and simulations.

  In traditional MOSFETs,  IRS has been  identified as  one of the most important scattering
  mechanisms \cite{And82}. Because of the strong confinement of carriers in  inversion layers that push them close to the semiconductor-oxide interface,
  the effective mobility of MOSFETs is limited by IRS at high electric fields or at high electron densities  \cite{Fang1968,Mazzoni1999}.

IRS in classical MOS type devices is treated via perturbation theory \cite{Ando1982}. It is represented by an IR-limited mobility depending on the root-mean-square (RMS) of the roughness amplitude and its correlation length \cite{Sakaki1934}. Early 1--D quantum device simulations of IRS in the NEGF formalism also used a perturbation treatment \cite{Klimeck1998}. In 3D nanowire devices, an explicit IR representation is feasible and could represent the device- to- device fluctuation. Recent theoretical studies have shown that mode-mixing due to IRS is reduced in NWFETs compared to conventional MOSFETs \cite{Wang2005,Buran2009,Poli2008} and the potential fluctuation throughout the NW dominates IRS in small cross-section NWs \cite{Poli2008}. However, these studies have been limited to a continuum representation of  rough surfaces with $\left<100\right>$ as transport direction and cross-sections larger than 3 $\times$ 3 nm$^2$. 

Atomistic full-band simulations offer a more realistic representation of rough surfaces  for any crystal directions. Atomistic simulations show that the ON-currents and the mobilities of NWFETs are reduced significantly by IRS when the diameter of NWs is below 3 nm\cite{Svi2007,Persson2008}. These simulations have been so far limited to ultra-narrow NWs without any representation of the SiO$_2$ layer, i.e. hard wall boundary conditions are applied to the silicon surface and the electrons do not penetrate into the oxide layer.

In this work, IRS is investigated in GAA $<$110$>$ NWFETs with different diameters (2 nm -- 4 nm) through an atomistic representation of the rough surface. Contrary to previous studies, the SiO$_2$ layers are modeled explicitly and taken into account not only in the Poisson equation, as in Ref. \cite{Luisier2007}, but also in the quantum transport calculation.
  The oxide layers are modeled in the virtual crystal approximation (VCA) where fictitious SiO$_2$ atoms are used. Disorder is introduced through a disordered spatial representation of Si and ``SiO$_2$'' atoms. The inclusion of the wavefunction penetration into SiO$_2$ allows us to investigate the ON-current and mobility reduction due to IRS in NWFETs and to compare simulation results to the experimental work in Ref. \cite{Suk2007}.

The paper is organized as follows: In Section \ref{sec:method}, the simulation approach is introduced with an emphasis on the  generation of rough surfaces in circular nanowires. In Section \ref{sec:results}, the simulation results are presented and discussed.

\section{\label{sec:method}Method}

\subsection{\label{subsec:roughness}Interface roughness model}

\begin{figure}
\subfigure[]{\label{fig:atomcross}\includegraphics[width=45mm]{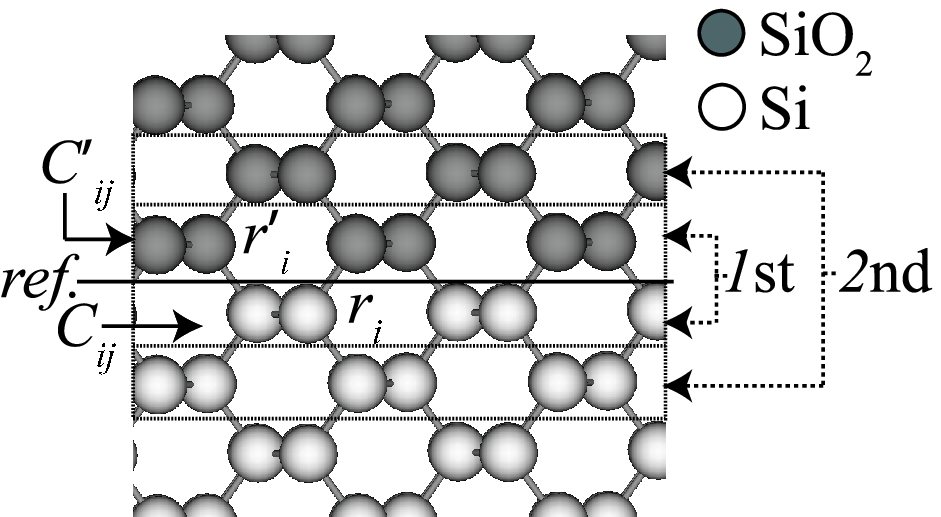}}
\subfigure[]{\label{fig:surface}\includegraphics[width=42mm]{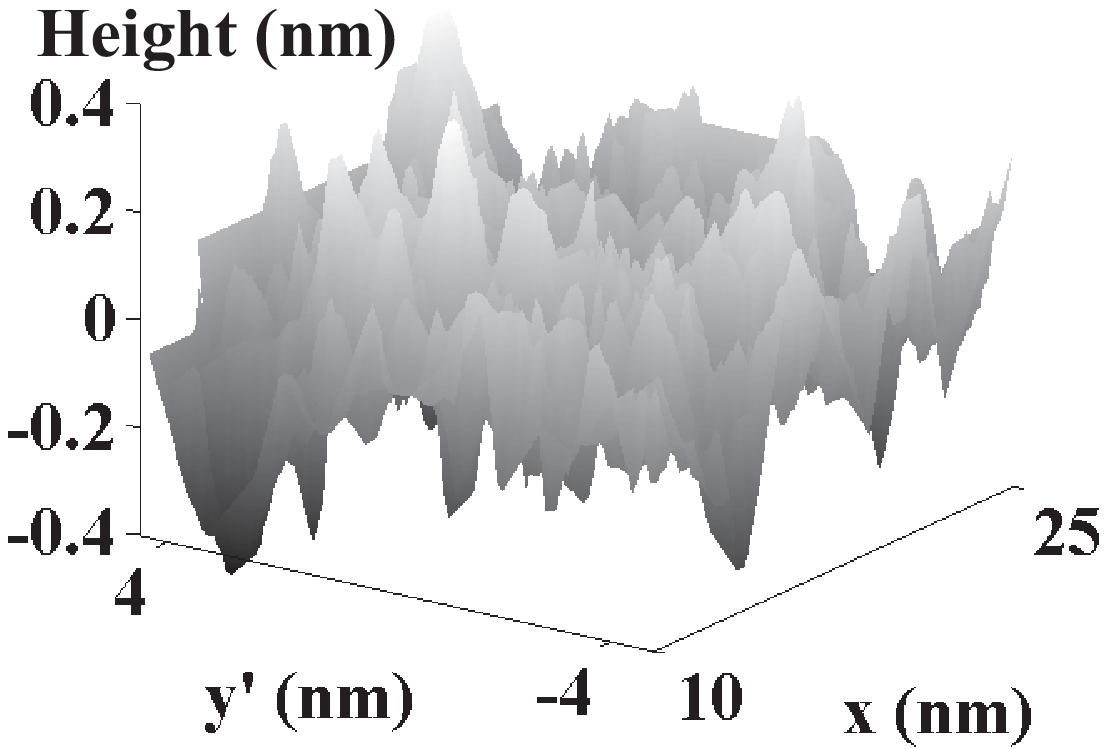}}
\caption{(a) Atomistic representation of the cross-section of a perfect nanowire along the $\left<110\right>$ direction. (ref.: reference plane, $C_{ij}$/$C'_{ij}$: ACVF for Si/SiO$_2$, $\bs{r}_i$/$\bs{r}'_i$: position vector of $i$-th Si/SiO$_2$ atom) (b) 2D surface profile generated for a nanowire of diameter $3$ nm with $\mathit \Delta_{\tx{m}}=0.14$ nm, $L_m=0.7$ nm for the NWFET structure presented in Fig. \ref{fig:NWFET} ($y'$ is a coordinate along the circumference direction.)}
\end{figure}

The statistical properties of the interface between Si and SiO$_2$ is characterized by the autocovariance function (ACVF)
\begin{align}\label{eq:autocov}
C_{ij}=\left< s\left( \bs{r}_i \right) s\left(\bs{r}_j\right) \right> = \mathit \Delta_{\tx{m}}^2e^{-\sqrt{2}\left|\bs{r}_i-\bs{r}_j\right|/L_m} \tx{ ,}
\end{align}
\noindent where $\mathit {\Delta_{\tx{m}}}$ is the root mean square (RMS) of the interface height and $L_m$ the correlation length of the rough surface.

 Here, $s(\bs{r}_i)$ is a random number representing the amplitude of the rough surface at position $\bs{r}_i$ where a Si atom is located and connected to SiO$_2$ atoms as shown in Fig. \ref{fig:atomcross}. The quantity $s(\bs{r}_i)$ can be defined as the distance between the position of the Si atom and the reference plane. This model has been developed from an experimental study\cite{Goodnick1985} where the measurement on the roughness amplitude is performed at an atomistic level. It is used in most of the simulation studies on IRS in NWFETs \cite{Wang2005,Luisier2007,Buran2009,Poli2008}. Notice that the $\sqrt{2}$ term in the numerator of the exponent in Eq. (\ref{eq:autocov}), which is not present in Ref. \cite{Luisier2007}, is included to correctly fit the model to the experiment as in Ref. \cite{Goodnick1985}.

From Eq. (\ref{eq:autocov}), one can calculate $s(\bs{r}_i)$ for each Si atom at the interface following the procedure in Ref. \cite{LuisierThesis}.
 Then, $s(\bs{r}'_i)$ for the SiO$_2$ atoms at the interface should be calculated by averaging the $s(\bs{r}_i)$ of Si atoms that are connected to a SiO$_2$ atom located at $\bs{r}'_i$ as shown in Fig. \ref{fig:atomcross}.
This introduces an error into the ACVF. For example, for a correlation length $L_m = 0.7$ nm, the error due to this approximation is approximately $1.88 \%$ only (see Appendix \ref{sec:appCij} for a more detailed derivation).

After $s(\bs{r}_i)$ and $s(\bs{r}'_i)$ are calculated for the atoms located at the first Si-SiO$_2$ interface layer as shown in Fig. \ref{fig:atomcross}, the $s(\cdot)$ for  atoms at the second layer are calculated in the same manner. Then, Si atoms are replaced by SiO$_2$ atoms and vice versa depending on the value of $s(\cdot)$ and a selection criterion. The criterion is described as follows: let $d(\bs{r}_i)$ be the distance from the atom considered to the reference plane located in between Si and SiO$_2$ atoms. Then, if $s(\bs{r}_i)>d(\bs{r}_i)$, the Si atom at $\bs{r}_i$ should be replaced by SiO$_2$ and if $s(\bs{r}'_i)<-d(\bs{r}'_i)$, the SiO$_2$ atom at $\bs{r}'_i$ should be replaced by a Si atom.

\begin{figure}
\centering
\includegraphics[width=80mm]{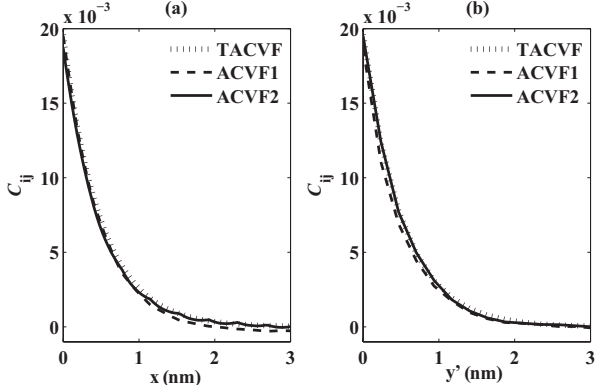}
\caption{(a) The average of ACVFs for 100 samples to $x$ direction, and (b) to the circumference direction ($y'$) are displayed. The halves of the graphs (a,b) to the negative direction are omitted because of their symmetry. All the calculations are done for NWs with diameter 3 nm and gate length 15 nm (see Fig. \ref{fig:NWFET}).}
\label{fig:acvf}
\end{figure}

The rough interface of a NW of diameter 3 nm and length 15 nm is projected to a two dimensional (2D) plane and plotted in Fig. \ref{fig:surface}. 
The rapid short-range fluctuation of the surface profile is the characteristic of {the chosen} exponential model rather than a Gaussian model \cite{Goodnick1985}.

At this point, it is necessary to verify that the rough interface profile generated with the algorithm explained above has the same statistical properties as the theoretical ACVF (TACVF) presented in Eq. (\ref{eq:autocov}). It is not clear in the literature \cite{Wang2005,Poli2008,Buran2009,Lenzi2008} whether the generated rough interfaces are correct representations of theoretical ACVF or not. Only in Ref. \cite{Khan2007}, the one dimensional (1D) ACVF of a single sample is compared to its theoretical value and it is shown that the first few coefficients of the ACVF from the lowest order agree with the TACVF.

\begin{figure}
\centering
\includegraphics[width=85mm]{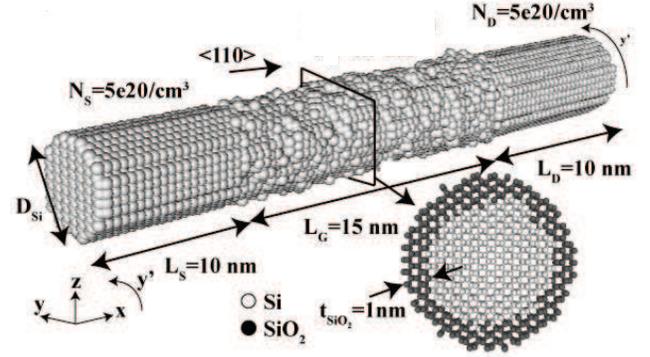}\caption{The simulated structure of NWFET: A circular NW with varying diameter ($D_{\tx{Si}}$=2, 2.5, 3, and 4 nm) is used. The default values of the other parameters are displayed in the figure.}
\label{fig:NWFET}	
\end{figure}

In this paper, 2D ACVFs of the generated rough interface are compared to the TACVF after averaging ACVFs of 100 samples. The ACVFs can be obtained either from $s(\bs{r}_i)$ (labeled ACVF1) or from the position of the rough interface atoms (labeled ACVF2). 
ACVF1 should agree with TACVF and ACVF2 with ACVF1 and consequently with TACVF. 
The following equation is used for 2D ACVF calculation \cite{Mar2004}
\begin{align}\label{eq:acvf}
C_{ij}={1 \over N_x N_y} \sum_{m=0}^{N_x-i-1} \sum_{n=0}^{N_y-j-1}(f_{m,n}-\bar f)(f_{m+i,n+j}-\bar f) \tx{,}
\end{align}
where $f_{m,n}$ is the amplitude of the rough surface perturbation projected onto the regular grids, and $\bar f$ is the average value of $f_{m,n}$. $N_x$ and $N_y$ are the numbers of grid points on the $x$ and $y'$ axis (see Fig. \ref{fig:NWFET} for a reference), respectively.

Since it is not easy to directly compare 2D ACVFs with each other, cutting sections of 2D ACVFs through the origin to $x$ and $y'$ direction are compared in Fig. \ref{fig:acvf}. The ACVF1 and the ACVF2 agree with TACVF for both $x$ and $y'$ direction when they are averaged over the entire rough NW samples. This validates the algorithm used for generation of rough interfaces in this work. The importance of the correct representation of interface roughness can be illustrated in the effect of the RMS value (=$\sqrt{\tx{ACVF}(0,0)}$) on the drain current from Fig. \ref{fig:Ionrms}.

\subsection{\label{subsec:SiO2model}SiO$_2$ model}

Previous attempts to model oxide materials in the tight-binding framework have been made by assuming a well-defined
crystallographic structure of SiO$_2$ such as $\beta$-quartz, tridymite \cite{staedele2001}, or $\beta$-cristobalite \cite{Sacconi2007}.  All these approaches have in common a $sp^3$ description of SiO$_2$ up to second nearest-neighbor interactions for the oxygen atoms. Their application is limited to one-dimensional Si-SiO$_2$-Si structures with a semiconductor-oxide interface perpendicular to the $<$100$>$ crystal axis only.

As an alternative and to go beyond one-dimensional $<$100$>$ transport we introduce a simplified, but multi-functional model to describe silicon dioxide. Fictitious SiO$_2$ atoms are considered and treated in the virtual crystal approximation. They are arranged as a diamond crystal with the same lattice constant $a_0=0.543$ nm as the Si atoms. A nearest-neighbor $sp^3$ tight-binding model is used to describe the electrical properties of the SiO$_2$. The band gap and conduction band offset are fitted to the experimental values with the relations $E_g=8.9$ eV and $\mathit \Delta E_{CB}=3.15$ eV with respect to Si \cite{Ludeke1999} as well as an electron effective mass $m_e=0.44 m_0$ \cite{Bersch2008}. A detailed description of the model is presented in Appendix \ref{sec:appsio2}.

\section{\label{sec:results}Results}

\begin{figure}[t]
\centering
\includegraphics[width=73mm]{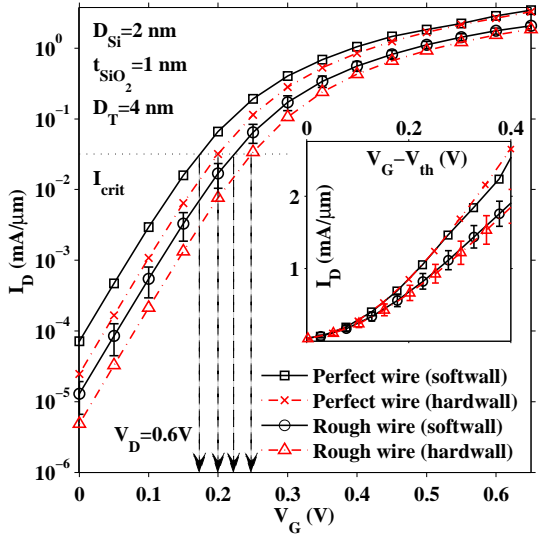}\caption{$I_D$ vs $V_{\tx{G}}$ for NWFETs of $D_{\tx{Si}}=2$ nm and total diameter $D_{\tx{T}}=D_{\tx{Si}}+t_{\tx{SiO}_2} \times 2=4$ nm. $I_{crit}(=D_{\tx{Si}}\times 10^{-7} \tx{A})$ is the critical current for the threshold voltage calculation. The gate work function for this simulation is chosen to be 3.95 eV. $V_{\tx{D}}$ is set to 0.6 V. The error bars indicate the standard deviation of the current in each bias point.}
\label{fig:Id}
\end{figure}

The NWFET structure considered in this work is illustrated in Fig. \ref{fig:NWFET}. $\mathit{\Delta}_{\tx{m}}=0.14 \tx{ nm}$ and $L_m=0.7 \tx{ nm}$ are used as default parameters which are experimentally measured in Ref. \cite{Goodnick1985} for a 2D Si/SiO$_2$ interface. Crystal orientation $\left<110\right>$ is selected as the transport direction because of its superior transport characteristics compared to other crystal orientations \cite{Phytos2008Elec} and the highest immunity to interface roughness scattering\cite{Luisier2007}. Two important parameters the diameter of the NW channel $D_{\tx{Si}}$ (depicted in Fig. \ref{fig:NWFET}) and $\mathit{\Delta}_{\tx{m}}$ are varied to study their impact on the NW performances. The diameter of silicon NW channel is varied from  2, 2.5, 3, to 4 nm and the RMS of the roughness amplitude is varied from 0.14, 0.2, to 0.3 nm. Strain is not considered in this work. The calculated $I_D$--$V_{\tx{G}}$ characteristics of perfect and rough NWFETs of 100 samples with a diameter of 2 nm are shown in Fig. \ref{fig:Id}.

Fig. \ref{fig:Id} presents two effects of IRS on $I_D$--$V_G$. The first effect is the positive shift of the threshold voltage and the second effect is the reduction of the ON-current. The positive shift of threshold voltage is ascribed to the fact that the diameter of rough NWs throughout the channel is reduced (on average) from the perfect softwall NW  due to rough interfaces\cite{Martinez2010} and that the density of states (DOS) in the energy spectrum is raised as discussed in Ref. \cite{Wang2005}. The amounts of threshold voltage shift on average are 49.2, 12.5, 8.1, and 10.0 mV for rough NWs with diameter 2, 2.5, 3, and 4 nm. The standard deviation $\sigma$ of the threshold voltage shift is 13.6, 10.1, 5.5, and 2.0 mV for each diameter from 2 to 4 nm.

The impact of IRS on the ON-current is measured by the reduction of the ON-current $\left(I_{\tx{perf}}-I_{\tx{scatt}}\right)/I_{\tx{perf}} \times 100$ where $I_{\tx{perf}}$ is the drain current for perfect NWs with softwall BC and $I_{\tx{scatt}}$ is the drain current for rough NWs. The ON-currents of rough NWs are reduced from perfect NWs by 20.5 \% ($D_{\tx{Si}}=2 \tx{ nm}$), 16.5 \% ($D_{\tx{Si}}=2.5 \tx{ nm}$), 9.0 \% ($D_{\tx{Si}}=3 \tx{ nm}$), and 3.8 \% ($D_{\tx{Si}}=4 \tx{ nm}$) due to IRS on average.

\begin{figure}[!b]
\centering\includegraphics[width=70mm]{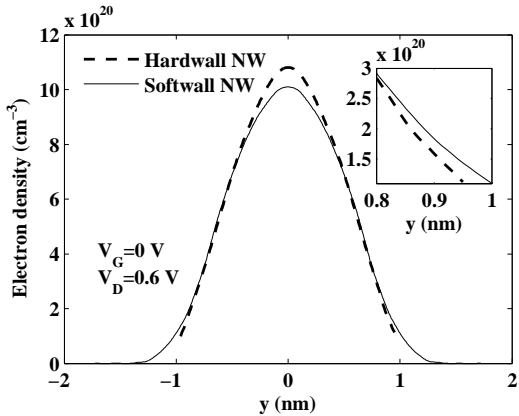}
\caption{Electron density along $y$-axis in the middle of the channel at $\left(z=0,x=L_{\tx{G}}/2\right)$ of a nanowire transistor with hardwall or softwall BC in the Si and SiO$_2$ interfaces. The inset describes the Si-SiO$_2$ interfaces around $y=1 \tx{ nm}$.}
\label{fig:effox}
\end{figure}

The effect of hardwall/softwall BC on the threshold voltage is noticeable as shown in Fig. \ref{fig:Id}. It can be explained from Fig. \ref{fig:effox} where the electron density is plotted through the center of the hardwall/softwall perfect NWs at $\left(z=0,x=L_{\tx{G}}/2\right)$. An increase of the carrier concentration near the interface between Si and SiO$_2$ layer in the softwall NW reduces the effective oxide thickness leading to the decrease of the threshold voltage.

The standard deviation of the threshold voltage due to IRS in the hardwall rough NWs (for 100 samples) for diameter 2 nm is calculated to be 14.3 mV, that is, 5\% increase compared to that in the softwall rough NWs. The effect of BC on the ON-current is even larger. The ON-current reduction in the hardwall NWs is 30\% which is 9.5\% higher compared to 20.5\% reduction in the softwall NWs. The standard deviation of the ON-current in the hardwall NWs is 0.179 mA/$\mu$m compared to 0.141 mA/$\mu$m in the softwall NWs, that is, 26.7\% increase. As the effective cross-section of the hardwall NWs is smaller than that of the softwall NWs, the fluctuation of the conduction band edge due to IRS is larger in the hardwall NWs and hence the calculation using the hardwall BC overestimates the effects of IRS.

\begin{figure}[t]
\centering
\subfigure{\includegraphics[width=63mm]{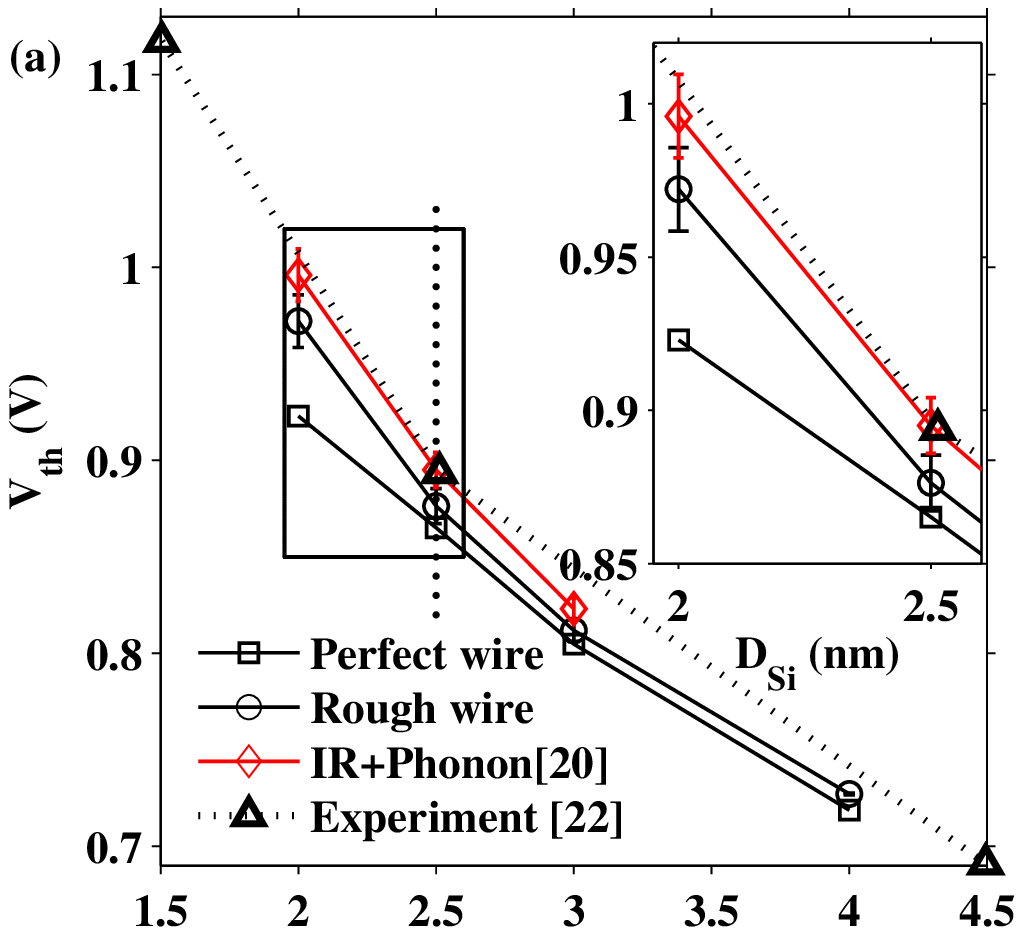}\label{fig:Vthdia}}
\subfigure{\includegraphics[width=63mm]{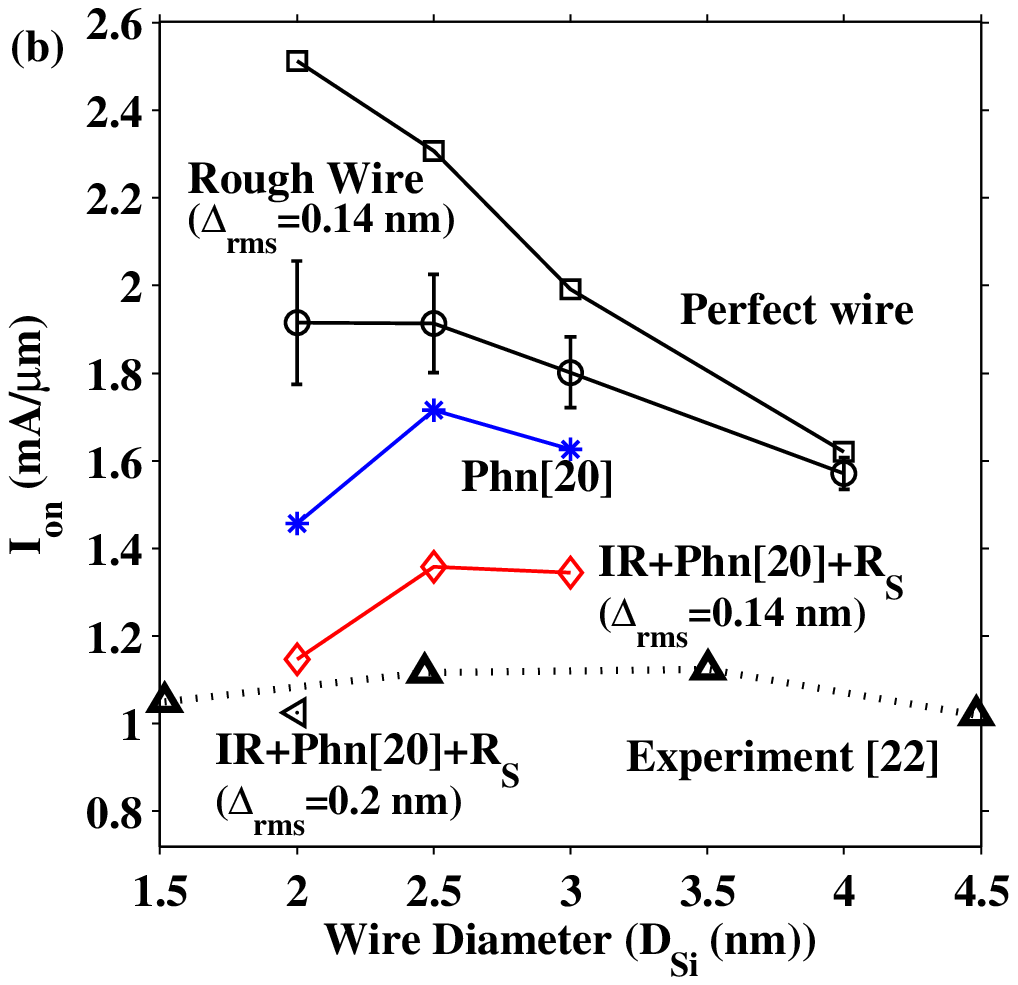}\label{fig:Iondia}}
\caption{(a) The threshold voltage and (b) the ON-current at $V_{\tx{G}}-V_{\tx{th}}=0.4 \tx{V}$ vs diameter for the perfect and rough NWFETs. The experimental ON-current and the source series resistance $R_S$ are extracted from Ref. \cite{Suk2007}. The results with phonon scattering are extracted from Ref. \cite{Luisier2009}. The inset in (a) magnifies the threshold voltage trend in the square box.}
\label{fig:IonVthdia}
\end{figure}

The threshold voltage and the ON-current for NWFETs with different diameters are plotted in Fig. \ref{fig:IonVthdia}. The threshold voltage is calculated using the constant current method \cite{Schroder2006,Wang2005} with a critical current $I_{crit}(=D_{\tx{Si}}\times 10^{-7} \tx{A})$. The threshold voltage shift due to phonon scattering is extracted from Ref. \cite{Luisier2009}. Because phonon scattering reduces the drain current for the whole range of the gate bias, the threshold voltage is also shifted to achieve the same critical current. The threshold voltages calculated for the perfect NW and the rough NWs have similar values and slopes when the diameter of the NWs is larger than 2.5 nm, and show similar trends as the experimental values extracted from Ref. \cite{Suk2007}. However, as the diameter decreases below 2.5 nm, the slope of the $V_{\tx{th}}$--$D_{\tx{Si}}$ curve for the perfect NW deviates from the experimental values. This implies that to get a consistent trend or quantitatively matched threshold voltage, one needs to include IRS in the transport calculation when the diameter of NW is scaled below 2.5 nm.

The drain current calculated at $V_{\tx{G}}-V_{\tx{th}}=0.4 \tx{V}$, i.e., the ON-current is shown in Fig. \ref{fig:Iondia}. The ON-current reduction due to IRS at a diameter smaller than 3 nm is significant and partly explains the decrease of the ON-current in the experimental results \cite{Suk2007}.

The ON-current with phonon scattering is calculated from the ballisticity in Ref. \cite{Luisier2009} of a very similar structure where only the doping density in the source/drain was different. In Ref. \cite{Luisier2009}, the ballistic current is calculated using the electrostatic potential extracted from the phonon scattering simulation to exclude the contribution of the phonon scattering in the source/drain extension region.

Since fully atomistic NEGF calculations are extremely computationally intensive, requiring thousands of CPUs for a single I-V, we cannot afford to perform statistical sampling with IRS. We therefore extract the effect of phonon scattering via the ballisticity obtained in Ref. \cite{Luisier2009}. However, the simulations in Ref. \cite{Luisier2009} are done for $\left<100\right>$-oriented NWs whereby our simulations are conducted for $\left<110\right>$-oriented NWs. From the fact that $\left<110\right>$-NWs have approximately 20\% larger ballisticity than $\left<100\right>$-NWs when phonon scattering is included \cite{Luisier2010}, the ballisticities of $\left<110\right>$-NWs are approximated from those of $\left<100\right>$-NWs. The relationship $1/T_{\tx{IR+Phn}}=1/T_{\tx{IR}}+1/T_{\tx{Phn}}-1$ is used where the transmission $T$ is calculated from the ballisticity factor $\beta$ using the relation $T_{\tx{IR/phn}}=2\beta_{\tx{IR/phn}}/(1+\beta_{\tx{IR/phn}})$.

\begin{figure}[t]
\centering
\includegraphics[width=85mm]{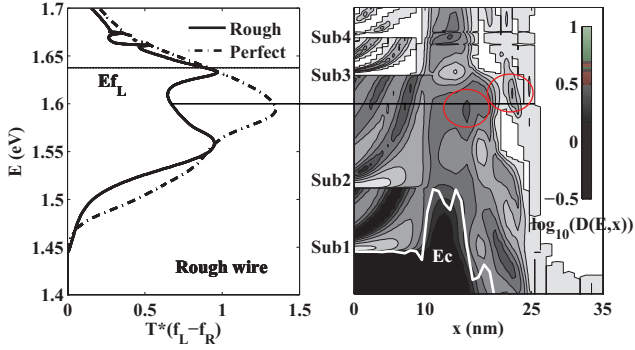}
\caption{(Left) the current spectra (=transmission $\times (f_{\tx{L}}-f_{\tx{R}}$)) with the Fermi level in the left contact $Ef_{\tx{L}}$ and (right) the density of states in a log scale across a rough nanowire FET resolved in energy $E$ and longitudinal coordinate $x$ at the ON-state ($V_{\tx{G}}-V_{\tx{TH}}\sim0.4$ V) overlapped with the conduction band edge $E_{\tx{C}}$.}
\label{fig:Tr_DOS}
\end{figure}

The ON-current reduction due to phonon scattering exceeds that of IRS, and the drain current including both phonon and IR scattering (with series resistance) are in accordance with the experimental results. Therefore, phonon scattering should also be included in the transport calculation to get a quantitative prediction for the ON-state characteristics of NWFETs.

Fig. \ref{fig:Tr_DOS} conveys detailed information about the effect of IRS on the current spectra and the DOS. IRS causes reflection of electrons and reduces the current spectra in the energy range where the peak is located in the perfect NW. A localization effect of the DOS, indicated by two circles, due to the fluctuation of the conduction band edge $E_{\tx{C}}$ in the channel causes this reduction of the current \cite{Svi2007,Martinez2010}. The current spectra and the DOS plot also shows that the relevant energy region of the electron transport includes subbands that are higher than the second subband. This supports that a full-band model is necessary to correctly understand and model these devices. 
\begin{figure}[t]
\centering
\subfigure{\includegraphics[width=75mm]{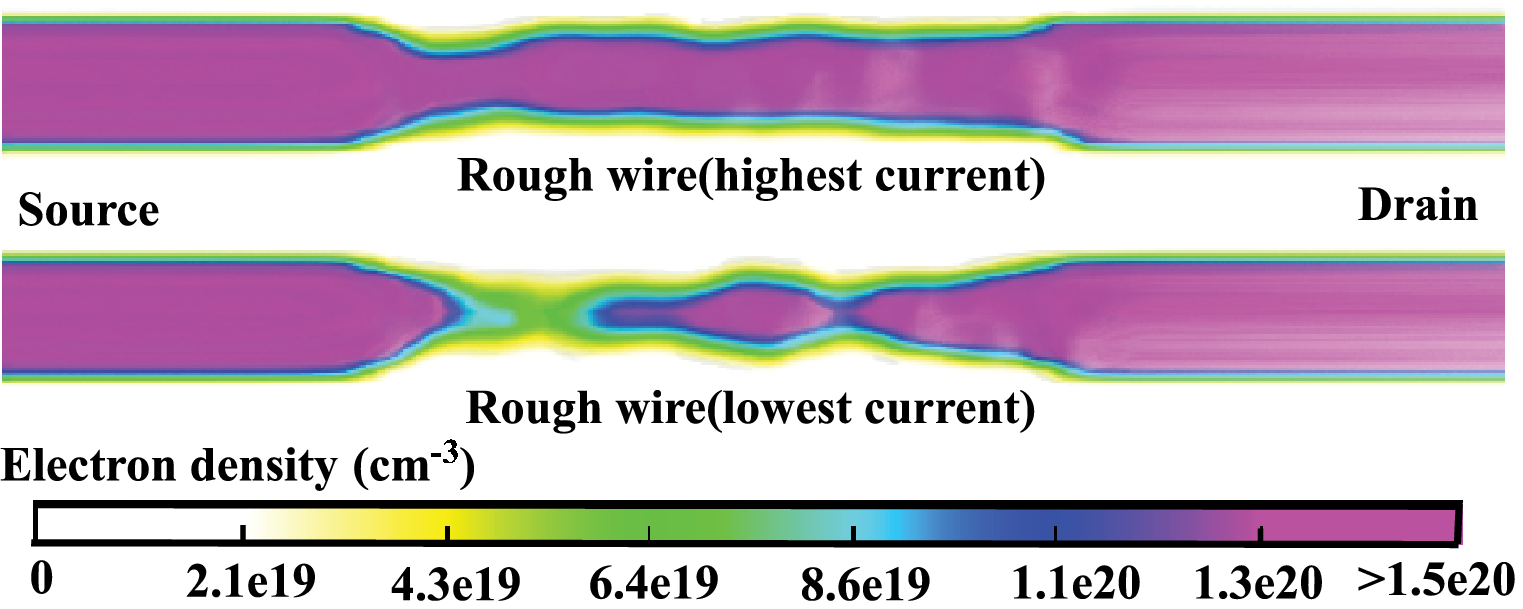}}
\subfigure{\includegraphics[width=75mm]{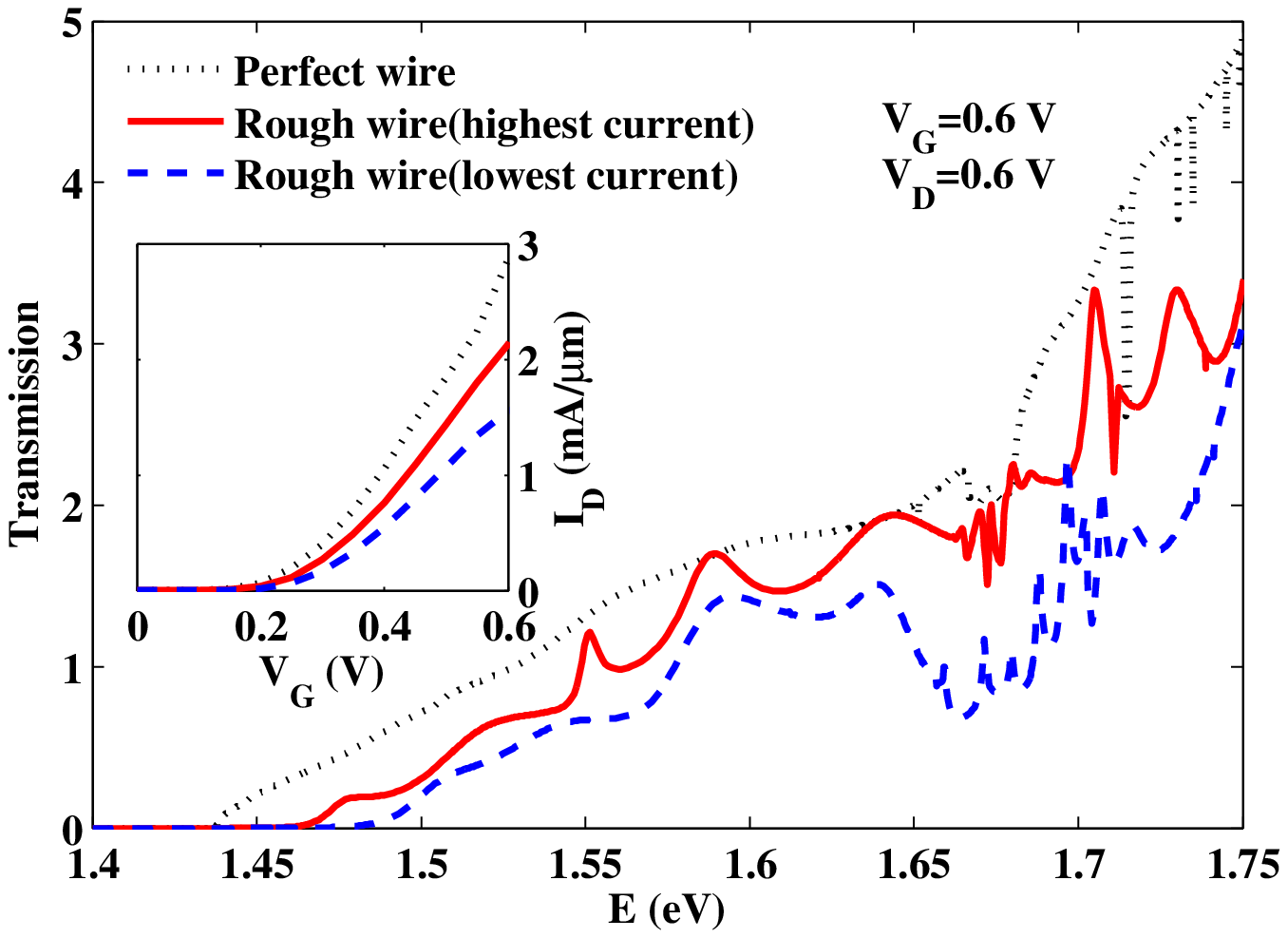}}
\caption{(Top) the electron density profile through the rough NW with higest/lowest drain current, (bottom) transmission through the perfect NW, the rough NW with the highest/lowest drain current at $V_{\tx{G}}=V_{\tx{D}}=0.6 V$, and (inset) $I_{\tx{D}}-V_{\tx{G}}$ plot.}
\label{fig:Trhigh}
\end{figure}

\begin{figure}[!b]
\centering
\includegraphics[width=48mm]{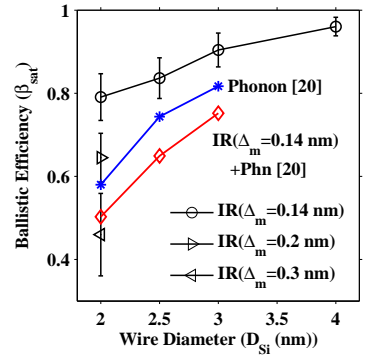}
\caption{The ballistic efficiency $\beta_{\tx{sat}}$ which is the ballistic ON-current divided by the average ON-current with IRS, the phonon scattering (extracted from Ref. \cite{Luisier2009}), or both the IR and the phonon scattering included.}\label{fig:TBvsdia}
\end{figure}
The transmission curves for the perfect NW, the rough NW with the lowest/highest drain current in Fig. \ref{fig:Trhigh} shed light on the impact of IRS on the electron transport characteristics of the nanowire transistor. The decrease in transmission due to IRS throughout the whole range of energy is observed. For the rough NW with the smallest drain current shows more localization effect in real space compared to the rough NW with the highest drain current, as can be seen from the top figure in Fig. \ref{fig:Trhigh}.

To further understand the importance of IR, the ballisticity is calculated for rough NWs with $D_{\tx{Si}}=2 \tx{nm}$ after $\mathit{\Delta}_{\tx{m}}$ is increased to 0.2 nm and 0.3 nm. It is then compared to the ballisticity including phonon scattering extracted from Ref. \cite{Luisier2009}. As shown in Fig. \ref{fig:TBvsdia}, the impact of IR on the ballisticity is smaller than that of phonon scattering for all diameters, but it becomes comparable to the phonon effects as $\mathit{\Delta}_{\tx{m}}$ increases to 0.2 nm and starts to exceed the phonon effects when $\mathit{\Delta}_{\tx{m}}=0.3 \tx{ nm}$.

\begin{figure}[t]
\centering
\subfigure{\includegraphics[width=43mm]{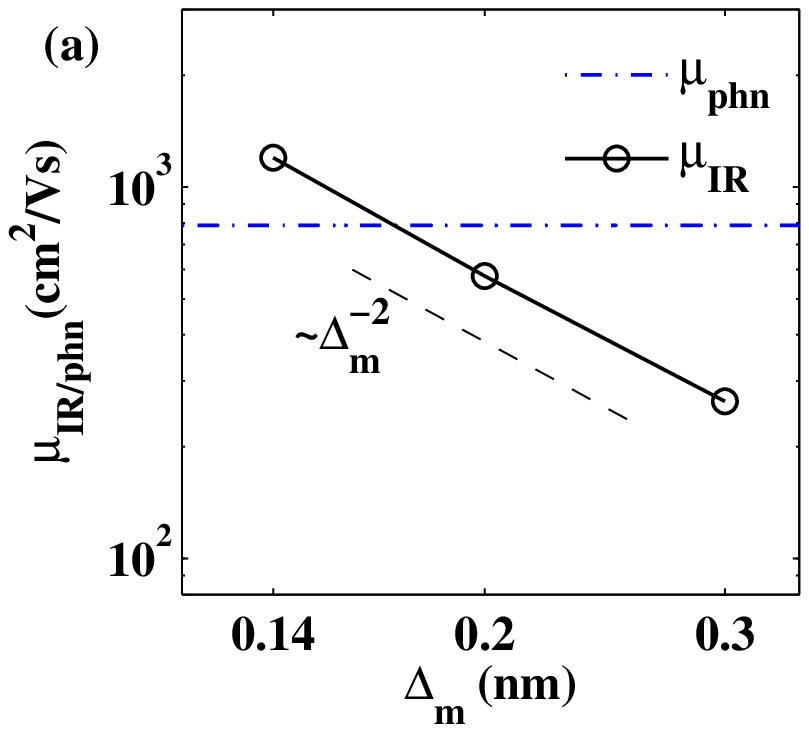}\label{fig:musrrms}}
\subfigure{\includegraphics[width=43mm]{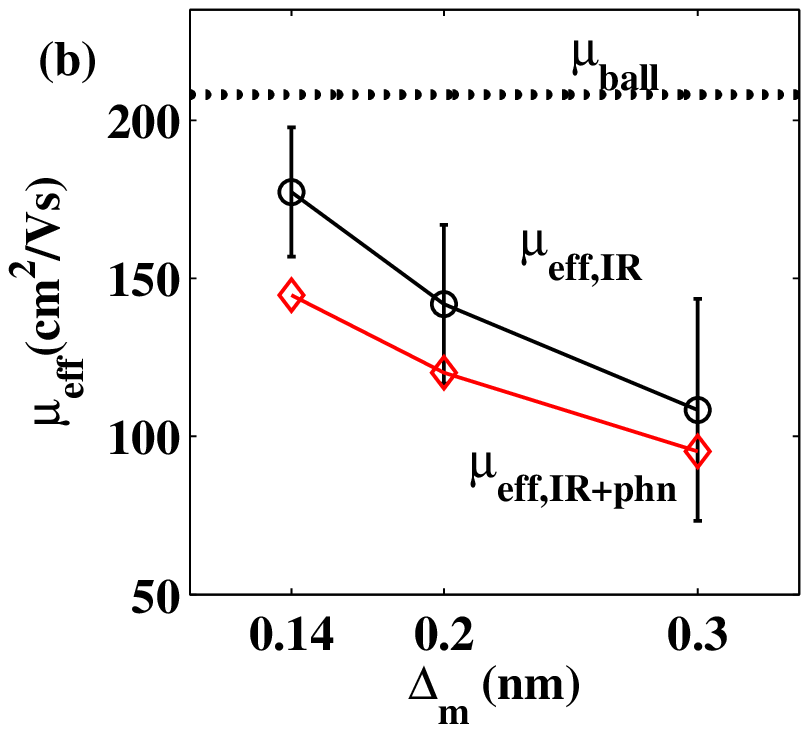}\label{fig:murms}}
\subfigure{\includegraphics[width=43mm]{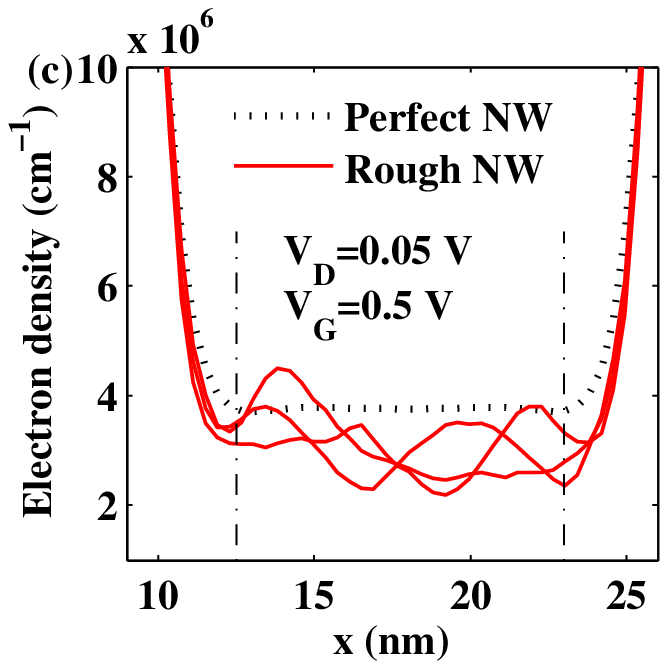}\label{fig:nx}}
\subfigure{\includegraphics[width=43mm]{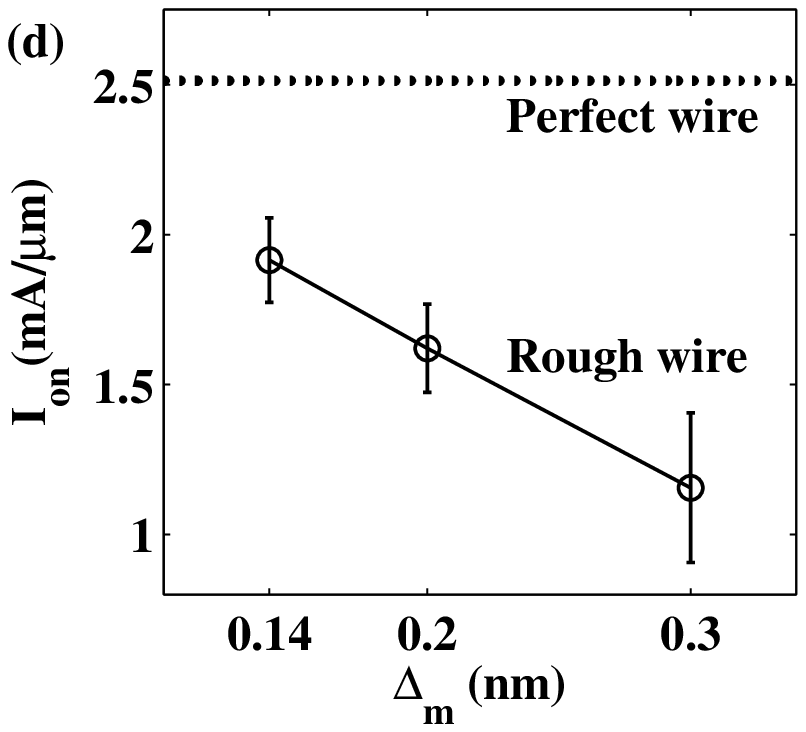}\label{fig:Ionrms}}
\caption{(a) The IR-limited mobility $\mu_{\tx{IR}}$ compared with the phonon-limited mobility $\mu_{\tx{phn}}$, (b) the total effective mobility including only IR or both the IR and the phonon scattering through Matthiessen's rule compared to the ballistic mobility $\mu_{\tx{ball}}$, (c) 1D electron density along the channel with the vertical bars defining the subsection of the channel for mobility calculation, and (d) the IR-limited ON-current compared with the ballistic current are displayed. All the figures are plotted for the nanowires with diameter $D_{\tx{Si}}$=2 nm. The electron density is set to $N_{\tx{s}}= 3.5\times10^6 \tx{cm}^{-1}$ except for (c) where the gate/drain voltage is fixed.}
\end{figure}

As the ON-state characteristics of a transistor are also related to their effective mobility \cite{Lund2001,Khaki2006}, the electron mobility reduction due to IRS at low drain bias should be understood in relation with the ON-current reduction at high drain bias. The low-field effective mobility in NWFETs is calculated from the electron density and the conductance using the expression\cite{Poli2008,Buran2009}
\begin{align}
\mu_{\tx{eff}}={ G_{\tx{lin}} L_{\tx{ch}} \over q N_{\tx{ch}}}
\end{align}
where $G_{\tx{lin}}$ is the conductance in the linear region i.e. at low drain bias ($V_{\tx{D}}=5$ mV in this work) and $N_{\tx{ch}}$ is calculated by integrating the electron density in the subsection of the channel under the gate where the electron density is nearly uniform \cite{Poli2008} as illustrated in Fig. \ref{fig:nx}.

\begin{figure}[t]
\centering
\subfigure{\includegraphics[width=43mm]{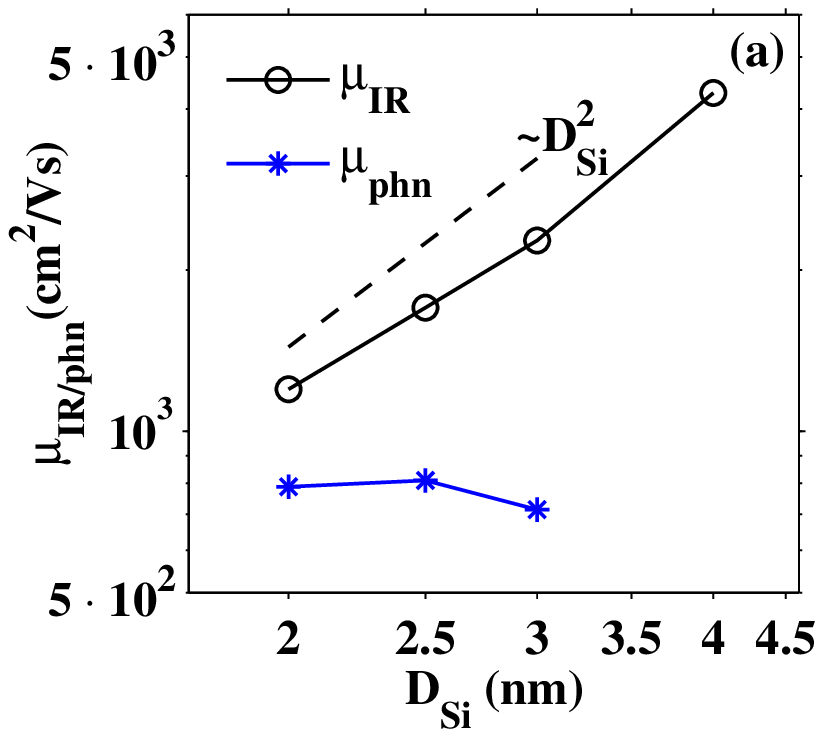}\label{fig:mulimdia}}
\subfigure{\includegraphics[width=43mm]{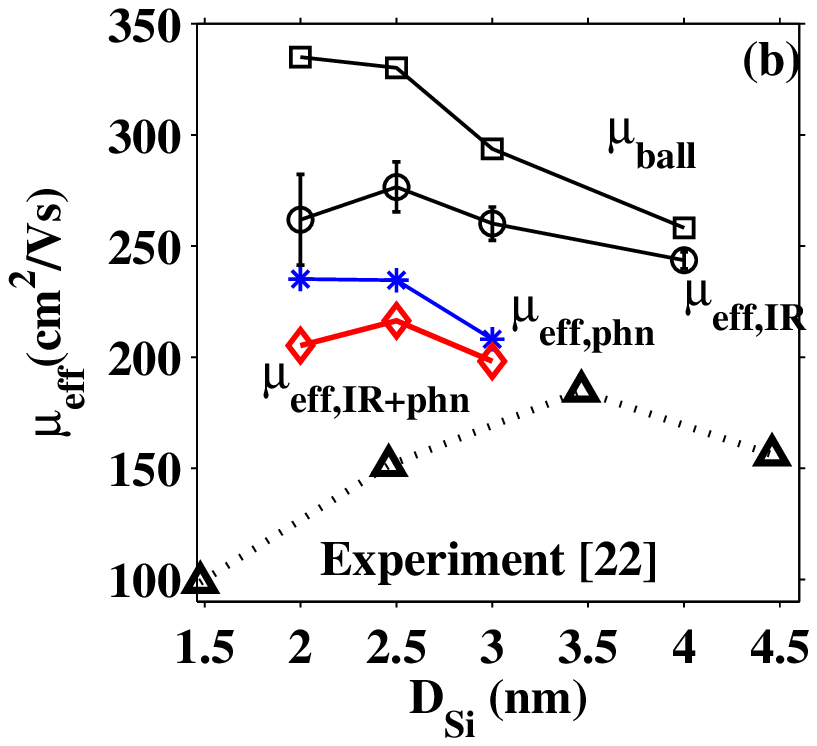}\label{fig:mueffdia}}
\caption{(a) The IR-limited mobility $\mu_{\tx{IR}}$ is calculated for the electron density $N_s= 3.5\times10^6 \tx{cm}^{-1}$ from the mean of the effective mobility including IRS $\mu_{\tx{eff,IR}}$ and the ballistic mobility $\mu_{\tx{ball}}$ through the Matthiessen's rule $\mu_{\tx{IR}}^{-1}=\mu_{\tx{eff,IR}}^{-1}-\mu_{\tx{ball}}^{-1}$. (b) The ballistic and effective mobility calculated for gate length $L_{\tx{G}}=30 \tx{nm}$ are compared to the experimental results which are extracted from Ref. \cite{Suk2007}. Effective channel length is taken into account by reducing the channel length below the gate length according to the experimental result \cite{Suk2007}.}
\label{fig:mudia}
\end{figure}

Assuming that the ballistic mobility is the effective mobility for a perfect NW, the IR-limited mobility $\mu_{\tx{IR}}$ can be calculated from the Matthiessen's rule, namely, $\mu_{\tx{IR}}^{-1}=\mu_{\tx{eff,IR}}^{-1}-\mu_{\tx{ball}}^{-1}$ where $\mu_{\tx{eff,IR}}$ is the effective mobility for rough NWs. The IR-limited mobility $\mu_{\tx{IR}}$ plotted in Fig. \ref{fig:mulimdia} is calculated from the mean of the effective mobility. It decreases as the diameter of the NW decreases. This is expected because IRS increases as the diameter decreases since the electrostatic potential fluctuates more in NWs with smaller diameter \cite{Jin2007} and it is consistent with the ON-current behavior in Fig. \ref{fig:Iondia}. The rate at which the average of $\mu_{\tx{IR}}$ decreases is approximately $\tx{D}_{\tx{Si}}^2$ which is much smaller than $\tx{D}_{\tx{Si}}^6$ for small electron density as described in Ref. \cite{Jin2007}.

To calculate the phonon-limited mobility, the resistance of the nanowire transistor is calculated for the same structure as shown in Fig. \ref{fig:NWFET}, but for several different channel length of nanowires (15, 22.5, and 30 nm). To remove the effect of the source and drain region, the phonon-limited mobility is calculated from the slope of the resistance\cite{Luisier2010}. 
The calculated phonon-limited mobility shown in Fig. \ref{fig:mulimdia} does not show a monotonic behavior with respect to diameter. It instead shows a peak at 2.5 nm and a decreasing trend as diameter increases to 3 nm. It can be explained by the argument that the higher subbands with a larger effective mass and scattering rates start to be populated with electrons as the diameter increases \cite{Zhang2010}.

Fig. \ref{fig:mueffdia} shows the effective mobility calculated for the rough and perfect NWs compared to experimental results. To compare the simulation results with experimental results, we consider the fact that the ballistic mobility increases proportionally to the channel length and that the effective channel length is smaller than the gate length as described in Ref. \cite{Suk2007}. The impact of IR on the effective mobility is still very small at diameter 4nm, but it increases as diameter decreases.

The discrepancy between the experimental results \cite{Suk2007} and the simulation results is noticeable even after phonon-limited mobility is included in the total effective mobility through Matthiessen's rule.
We attribute this difference to the different electron density at which the effective mobility is calculated in the experiment and our simulation. 
In the experiment, supply voltage $V_{\tx{DD}}$ used is higher than that used in our simulation so that the effective mobility at the ON-state in the experiment is smaller than that in the simulation. In addition, the surface quality of experimentally fabricated NWs may be different than that of the 2D plane surface which is adopted for our simulation. Nonetheless, IR scattering is crucial to understand the decreasing effective mobility in NWs of smaller diameter. 

Fig. \ref{fig:musrrms} describes the importance of the quality of the interfaces on the electron mobility in NWFETs for different $\mathit{\Delta}_{\tx{m}}$. The mean of the IR-limited mobility decreases with $\sim\mathit{\Delta}_{\tx{m}}^{-2}$ and it becomes comparable to the phonon limited mobility when $\mathit{\Delta}_{\tx{m}}= 0.2 \tx{ nm}$. This indicates that poor surface quality can reduce the electron mobility significantly.

When $\mathit{\Delta}_{\tx{m}}$ increases, not only does the effective mobility decrease on average as in Fig. \ref{fig:murms}, but also the standard deviation of the effective mobility increases. The standard deviations of the effective mobility with respect to the means of the effective mobility are given in percentages of 11.5 \% ($\mathit{\Delta}_{\tx{m}}$=0.14 nm), 17.7 \% ($\mathit{\Delta}_{\tx{m}}$=0.2 nm), and 32.4 \% ($\mathit{\Delta}_{\tx{m}}$=0.3 nm). The standard deviations of the ON-current in Fig. \ref{fig:Ionrms} are, in percentages, 7.37 \% ($\mathit{\Delta}_{\tx{m}}$=0.14 nm), 9.09 \% ($\mathit{\Delta}_{\tx{m}}$=0.2 nm), and 21.6 \% ($\mathit{\Delta}_{\tx{m}}$=0.3 nm) from the mean of ON-current. The variability of the ON-current due to IRS is smaller than that of the effective mobility. In conclusion, the quality of the interface between Si and SiO$_2$ is important in reducing the variability of the ON-current and the effective mobility in NWFETs.

\section{\label{sec:con}Conclusion}

The effect of interface roughness between Si and SiO$_2$ on the performance of NWFETs has been studied through an atomistic 3D full-band simulation. The SiO$_2$ layer is modeled through VCA and its impact on the drain current is found to be significant. IRS is important in determining the threshold voltage of NWFETs. The experimentally observed \cite{Suk2007} trend of nonlinear threshold voltage slope below 2.5nm is captured by the simulation. The significant reduction of the ON-current is observed in NWs with diameter 2 nm with more than 20 \% reduction.

Though the phonon scattering is found to be an important scattering mechanism in NWFETs, IRS cannot be ignored. IRS has a significant effect on both the ON-current and the effective mobility specially when the diameter decreases down to 2 nm. As the effect of IRS depends on the surface quality in NWFETs, the interface roughness scattering is a crucial factor to predict the device performance of NWFETs.

\appendices
\section{\label{sec:appCij}Calculation of Autocovariance function for SiO$_2$}
\begin{table}[!b]
\centering
\begin{tabular}{|c|c|c|c|c|c|c|}
\hline
&$E_s$&$E_p$&$V_{ss\sigma}$&$V_{sp\sigma}$&$V_{pp\sigma}$&$V_{pp\pi}$\\
\hline
SiO$_2$&-4.8&1.83&-2.27&3.9265&5.4655&-0.3140\\
\hline
\end{tabular}
\caption{sp$^3$ Tight-binding parameters for the VCA SiO$_2$ model.}
\label{tab:tb_par}
\end{table}
The random variables describing the rough SiO$_2$ surface, denoted as $s(\bs{r}'_i)$, must be calculated from the Si atoms because they are inter-correlated with each other. Let us assume that one SiO$_2$ atom is connected to two Si atoms whose random variables are $s(\bs{r}_i)$ and $s(\bs{r}_{i+1})$. Then we approximate $s(\bs{r}'_i)$ by averaging the two random variables as
\begin{equation}
s(\bs{r}'_i)={{s(\bs{r}_i)+s(\bs{r}_{i+1})}\over 2}\tx{.}
\end{equation}

\begin{figure}[!b]
\centerline{\includegraphics[width=1.00\linewidth]{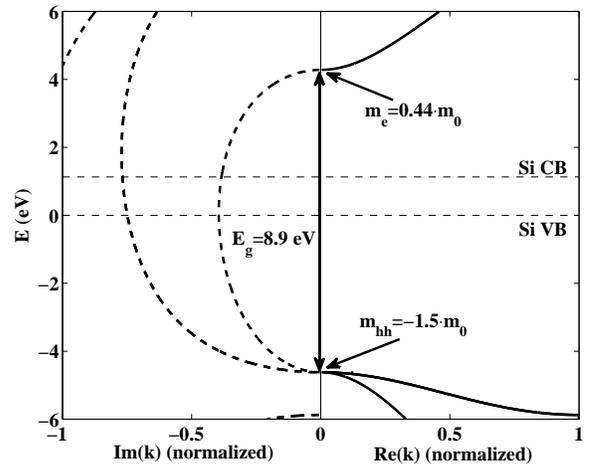}}
\caption{Bandstructure of the fictitious VCA diamond-like SiO$_2$
  material at $k_y$=$k_z$=0. Real and complex bands are shown as well
  as the silicon conduction and valence band edges, the electron and
  heavy-hole effective masses of the oxide, and its band gap.}
\label{bandstructure}
\end{figure}

This approximation induces an error in the autocovariance of the SiO$_2$ atom ($C_{ij}'$) as shown in the following calculation:
\begin{align}\label{eq:acvfsio2}
C_{ij}'&=\left<s(\bs{r}'_i)s(\bs{r}'_j)\right>\nonumber \\
&=\left<\left[{s(\bs{r}_i)+s(\bs{r}_{i+1})}\over 2\right]\left[{s(\bs{r}_j)+s(\bs{r}_{j+1})}\over 2\right]\right>\nonumber\\
&\cong\mathit \Delta_{\tx{m}}^2 e^{-\sqrt 2\left|\bs{r}_i-\bs{r}_j\right|/L_m}\left(2+e^{\sqrt 2\mathit\Delta r / L_m}+e^{-{\sqrt 2\mathit\Delta r / L_m}}\right)\nonumber\\
&=C_{ij}\left[{1+\cosh \left({\sqrt{2}\mathit{\Delta} r/L_m}\right) \over 2}\right] \tx{,}
\end{align}
where Eq. (\ref{eq:autocov}) and the relationship $\mathit{\Delta r}=\bs{r}_{i+1}-\bs{r}_i\cong\bs{r}_{j+1}-\bs{r}_j$ are used.
Then, Eq. (\ref{eq:acvfsio2}) can be further approximated with a Taylor expansion up to the second order as
\begin{align}
C_{ij}'\cong C_{ij}\left[1+{x^2 \over 4}\right]
\end{align}
with $ x = \sqrt{2} \mathit{\Delta r} / L_m$.
If $\mathit{\Delta r}\approx 0.1358$ nm and $L_m = 0.7$ nm, the difference between $C_{ij}$ and $C'_{ij}$ due to this approximation is $x^2/4= \mathit{\Delta r}^2 / (2L_m^2)\approx1.88 \%$.

\section{\label{sec:appsio2}SiO2 model and results}

The list of the $sp^3$ tight-binding parameters for SiO$_2$ is given in
Table~\ref{tab:tb_par} and the bandstructure, including the imaginary band
 linking the lowest conduction band to the light-hole band at $\Gamma$
 in Fig.~\ref{bandstructure}. The light-hole effective
mass is strongly correlated to the electron effective mass in the $sp^3$
tight-binding model ($m_{lh}$=-0.3438$m_0$). The
heavy-hole mass does not affect the SiO$_2$ imaginary band linking
the light-hole valence band and ``lowest'' conduction band at $\Gamma$.
It is this latter imaginary band which largely determines the wavefunction
decay constant in the oxide gap. Therefore, it is arbitrarily set to $m_{hh}$=-1.5$m_0$.

\bibliographystyle{IEEEtran}
\bibliography{IEEEabrv,Biblio_IRS_Kim}

\end{document}